\documentclass[useAMS,usenatbib]{mn2e}

\usepackage[authoryear]{natbib}
\usepackage{rotating}
\usepackage{lscape}
\usepackage{acronym}
\usepackage{subfigure}
\bibpunct{(}{)}{;}{a}{}{,}
\usepackage{graphicx}
\usepackage{epsf}
\usepackage{lscape}
\usepackage{color}
\usepackage{graphics}

\newcommand{\D}{$^\circ$}

\newcommand{\HII}{H\,{\sc ii}}

\newcommand{\SII}{[S\,{\sc ii}]}
\newcommand{\OI}{[O\,{\sc i}]}
\newcommand{\OII}{[O\,{\sc ii}]}
\newcommand{\OIII}{[O\,{\sc iii}]}
\newcommand{\NII}{[N\,{\sc ii}]}

\newcommand{\NeIII}{[Ne\,{\sc iii}]}
\newcommand{\HA}{H{$\alpha$}}
\newcommand{\AIII}{[Ar\,{\sc iii}]}

\newcommand{\HB}{H{$\beta$}}

\newcommand{\HG}{H{$\gamma$}}
\newcommand{\HE}{[He\,{\sc i}]}
\newcommand{\Halpha}{H$\alpha$}
\def\p0{\phantom{0}}
\def\arcmin{\hbox{$^{\prime}$}}
\def\arcsec{\hbox{$^{\prime\prime}$}}
\title[A multi-frequency study of SNR B0513--692]{Multi-frequency Study of the LMC Supernova Remnant (SNR) B0513--692 and New SNR Candidate J051327--6911}
\author[Boji\v{c}i\'c et al.]
  {I.S.~Boji\v{c}i\'c,$^1$
  M.D.~Filipovi\'c,$^{2}$ Q.A.~Parker,$^{1,3}$ J.L.~Payne,$^4$
  P.A.~Jones,$^5$
  \newauthor W.~Reid,$^1$ A.~Kawamura,$^6$ Y.~Fukui,$^6$\\
  $^1$Department of Physics, Macquarie University, Sydney, NSW 2109, Australia\\
  $^2$University of Western Sydney, Locked Bag 1797, Penrith South DC, NSW 1797, Australia\\
  $^3$Anglo-Australian Observatory, P.O. Box 296, Epping, NSW 1710, Australia\\
  $^4$Centre for Astronomy, James Cook University, Townsville, Queensland 4811, Australia\\
  $^5$Australia Telescope National Facility, CSIRO, P.O. Box76, Epping, NSW 1710, Australia\\
  $^6$Department of Astrophysics, Nagoya University, Furocho, Chikusaku, Nagoya 464--8602, Japan\\
}

\date{Accepted 2006 December 31. Received 2006 December 31}
\pagerange{\pageref{firstpage}--\pageref{lastpage}}
\pubyear{2006}

\begin{document}

\maketitle

\label{firstpage}

\begin{abstract}
We present a new multi-wavelength study of supernova remnant (SNR)
B0513--692 in the Large Magellanic Cloud (LMC). The remnant also has
a strong, superposed, essentially unresolved, but unrelated radio
source at its north-western edge, J051324--691049. This is
identified as a likely compact \HII\ region based on related optical
imaging and spectroscopy. We use the Australia Telescope Compact
Array (ATCA) at 4790 and 8640~MHz ($\lambda\simeq$~6~cm and
$\lambda\simeq$~3.5~cm) to determine the large scale morphology,
spectral index and polarization characteristics of B0513--692 for
the first time. We detect a strongly polarized region (49\%) in the
remnant's southern edge ($\lambda\simeq$~6~cm). Interestingly we
also detect a small ($\sim40$~arcsec) moderately bright, but distinct
optical, circular shell in our \HA\ imagery which is adjacent to the
compact \HII\ region and just within the borders of the NE edge of
B0513--692. We suggest this is a separate new SNR candidate based on
its apparently distinct character in terms of optical morphology in
3 imaged emission lines and indicative SNR optical spectroscopy
(including enhanced optical \SII\ emission relative to \HA).

\end{abstract}
\begin{keywords}
ISM: supernova remnants -- ISM: \HII\ regions -- galaxies:
Magellanic Clouds -- radio continuum: galaxies: B0513--692
(J051315--691219); N\,112: J051324--691049: J051327--6911: SNRs:
\HII\ regions.
\end{keywords}
\section{Introduction}

The Large Magellanic Cloud (LMC) offers a unique opportunity for
studying supernovae and supernova remnants (SNRs) in varied
environments. All objects within the LMC can be considered
essentially at the same distance of $\sim~51.5~\mathrm{kpc}$ (e.g.
\citealt{Feast1999}; \citealt{Nelson2000}) and is essentially a thin
($\sim500$~parsec) disk inclined at only 35~degrees to our line of
sight \citep{MarCio2001}. This permits accurate nebular and physical
parameters to be determined for LMC SNRs. Furthermore, the LMC is
close enough to resolve the structure and morphology of individual
extended objects contained within it, such as SNRs. These studies
are enhanced by the small uncertainty in absorption towards the LMC
(e.g. \citealt{KalJac1990}).

The best approach to identify SNR candidates is through the use of
multi-wavelength surveys. Most commonly, these surveys include the
\mbox{X-ray}, radio and optical domains, although SNRs also produce
a large number of emission lines in the ultraviolet (UV) and
infrared (IR) wavelength regimes. In the radio, techniques to
distinguish between SNRs and the other nebula types are based on
their strongly polarized and non-thermal emission. SNRs have an
average radio spectral index (defined as $\mathrm{S}_\nu \propto
\nu^{\alpha}$) of $\alpha=-0.5$ \citep{IngKit1990} compared to \HII\
regions which have much flatter spectra of $\alpha\sim$~0. In the
optical, narrowband imaging at H$\alpha$, \SII\ and \OIII\
wavelengths can be a useful discriminant as the morphological
details, such as spherical (shell) symmetry and filamentary
structure is helpful in distinguishing SNRs from other sources such
as \HII\ regions and superbubbles, especially as a function of the
imaged line. Comparison of their spectral emission-line signatures
from spectroscopy allows for more unequivocal identification of SNRs
via specific, diagnostic line ratios, particularly \SII\ relative to
\HA\ (e.g. \citealt{Fesen1985}).

The first SNR radio identification in the LMC was made using the
Parkes radio telescope at 1410~MHz with a resolution of about
14\arcmin\ \citep{Math1964} and the most extensive work in identifying
Magellanic Clouds SNRs was conducted by \citeauthor{Math1983}
(\citeyear{Math1983}, \citeyear{Math1984}, \citeyear{Math1985}).
Recently, we have identified 76 SNRs within the LMC of which 32 are new
SNR candidates (Filipovi\'c et al., in prep). Currently, the identifications
are based solely on each object's morphological structure from our
powerful new radio mosaic images of the LMC so additional corroborating data
such as presented here is extremely useful.

In this paper we investigate the multi-wavelength properties of one
of the LMC remnants recovered in our new survey: SNR~B0513--692. We do
this by bringing together new and archival radio, X-ray and optical imagery,
together with new optical spectroscopy and a fresh assessment of the
surrounding environment.

SNR~B0513--692 (J051315--691219) was first confirmed as a SNR by
\cite{Math1985} who studied its morphology, non-thermal radio
spectra and filamentary \Halpha\ emission.

In the radio band, they used detections from \cite{MilTur1984} with
the Molonglo Observatory Synthesis Telescope (MOST) at 843~MHz,
and from \cite{Cllimi1976} with Molonglo radio data at 408~MHz, to
examine the radio morphology and to determine a radio spectral
index of $\alpha=$-0.5. An initial \mbox{X-ray} detection of the
remnant candidate was based on data obtained from the archives of
the Columbia--Einstein Observatory survey of the LMC
\citep{Long1981}. In the optical they used the IPCS on the
3.9~m Anglo--Australian Telescope (AAT) coupled with narrow-band
interference filters centred on key optical emission lines to image a
$3.5\times3.5$~arcmin region of the remnant. In particular
they found the \SII/\HA\ point-to-point ratio of this object
was typically $\sim0.6$. Such a high value is a classic signature of
a likely SNR \citep{Fesen1985}. Their narrow-band observations showed
the filamentary \HA\ emission following the overall radio structure.
They also found an intense, compact emission component at the
North-East of the remnant which corresponds to a compact radio
source also noted. They identified this source as a likely,
unrelated, \HII\ region.

Here, we present a more detailed multi-radio frequency,
multi-wavelength study of this LMC SNR and its immediate environs
in order to provide a better picture of this little studied
object and to shed light on the nature of the
compact radio source on the North-Eastern edge and its optical
counterparts. We now propose that there are, in fact, two
additional, separate sources on the NE boundary of B0513--692. One
is confirmed as the previously catalogued compact \HII\ region DEM
L109, but the other appears to be a small, unrelated, compact new
SNR candidate which we designate J051237--6911, overlapping both the
\HII\ region and the larger remnant B0513--692.

\section{Compiled source Data for B0513--692: New and archival}

The new radio-continuum observations of SNR B0513--692 were
performed in 'snap-shot' mode (integration time was about
0.7~hours at each frequency) using the Australia Telescope Compact
Array (ATCA) at 4790~MHz and 8640~MHz on 6 April 1997. We used the
375-m array configuration which includes 10 baselines between the
first and fifth antenna and an additional five (5) spacings when a
6-km antenna was used for higher resolution imaging. Our
calibration sources included PKS~B1934--638 (primary flux
calibrator) and PKS~B0530--727 (phase calibrator).

\begin{figure}
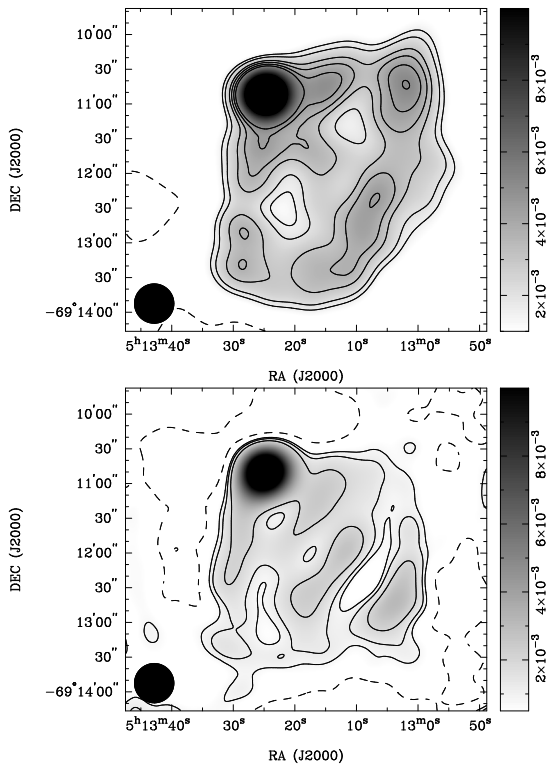

\begin{center}
\includegraphics[scale=0.38, angle=0]{int6.eps}
\includegraphics[scale=0.38, angle=0]{int3.eps}
  \caption{Total intensity ATCA images of SNR~B0513--692 at 4790~MHz (top)
and 8640~MHz (bottom), overlaid with contours giving the
associated flux radio level. For the 4790~MHz image, contours
include -1, 3, 4, 6, 7.3, 8.8, 10.3 and 20$\sigma$
($\sigma=0.5\textrm{ mJy Beam}^{-1}$) and for the 8640~MHz image
they are -1, 2, 3 and 5$\sigma$ ($\sigma=0.4\textrm{ mJy
Beam}^{-1}$). The synthesized beam of the both ATCA observations
are 34$\times$34~arcsec (lower left corner of each
image). Note the strong compact radio source at the NE edge of the
remnant.
 \label{Fig1}}
\end{center}
\end{figure}

Data reduction was performed using the {\sc miriad} software package
\citep{SauKil2003}. Radio-continuum images of these observations are
shown in Fig.~\ref{Fig1}. These images were formed using
multi-frequency synthesis \citep{SauWie1994} and natural weighting.
They were deconvolved using the {\sc clean} and {\sc restor}
algorithms with primary beam correction applied using the {\sc
linmos} task. A similar procedure was used for both \textit{U} and
\textit{Q} Stokes parameter maps. Because of the low dynamic range
(signal to noise ratio between the source flux and $3\sigma$ noise
level) self-calibration could not be applied.

A resolution of 34~arcsec was achieved at 6~cm
(against 43~arcsec from \cite{MilTur1984}).
Additionally, our new radio images are compared with the LMC radio
maps at 3 and 6~cm from \cite{Dickel}. Better sensitivity is achieved,
especially at 3~cm, because of the longer integration time. These
new data allows us to better delineate the overall shell structure
and, furthermore to provide solid detection of polarized radio
emission from this SNR.

\begin{table}
\begin{center}
\caption{Summary of the ATCA radio-continuum imaging parameters of
the region -- SNR~B0513--692.\label{Tab1}}
\small{\begin{tabular}{ccc} \hline \noalign{\smallskip}
Freq. & FWHM                      & FWHM                       \\
(MHz) &  (all ant.)           &  (6-km baseline excluded)  \\
\hline \noalign{\smallskip}
4790  & 2\arcsec$\times$~2\arcsec  & 34\arcsec$\times$~34\arcsec  \\
8640  & 1\arcsec$\times$~1\arcsec  & 19\arcsec$\times$~19\arcsec  \\
\hline
\end{tabular}}
\end{center}
\end{table}

The strong, compact radio source,
J051324--691049, embedded within the North-East side of the SNR
complicates the radio picture. In order to attempt to resolve
this object we used all telescope baselines (i.e. all correlations
with the sixth antenna) and a Gaussian restoring beam to create a
high-resolution image with a resolution of 1~arcsec at
3~cm. Unfortunately, due to the nature of the radio
interferometric technique, incomplete uv coverage between the 375~m
and 6~km ATCA baselines could result in over-resolving or breaking
up of the source, if it is extended, into multiple point-like components.
We didn't find any artifacts of possible distortion, and we note that
the source is either unresolved at 1~arcsec or is no larger than
19~arcsec, which is the highest resolution of the 375~m baseline at
8640~MHz. We re-analysed the ATCA data (but omitting the 6-km
baseline) to yield a lower resolution but higher sensitivity image to
enable a better study of the extended, diffuse nature of
SNR~B0513--692. Table~\ref{Tab1} lists the basic radio imaging
parameters. To enhance our study of the extended remnant
SNR~B0513--692 and the unrelated, embedded, compact source
J051324--691049, we also examined existing radio data from the
Sydney University Molonglo Sky Survey (SUMSS) at 843~MHz
\citep{Bock1999} and a new ATCA mosaic image at 1377~MHz
(Staveley-Smith et al., in prep.).

We also analysed new, narrow--band optical data of the SNR extracted
from a deep, high resolution \HA\ map  created by the digital
combination (median stacking) of $12\times2$~hour, narrow-band, \HA\
UK Schmidt Telescope (UKST) exposures of the central 25 square
degrees of the LMC (Reid \& Parker 2006a,b), also section 3.2.4). This was
supplemented by independent \HA, \SII\ and \OIII\ images from the
Magellanic Cloud Emission Line Survey (MCELS) survey
(http://www.ctio.noao.edu/mcels/)\footnote{The relevant
$6\times6$~arcmin fits images of the region were kindly made
available to us by C.Smith et al. prior to general release.}. These data have
proven crucial in helping to unravel the nature of the compact NE
components (see later).

The CO map of the region comes from the NANTEN survey of
the LMC \citep{Fukui1999,Fukui2001}. Infrared images at 12, 25, 60
and 100~$\mu$m were also obtained from the Infrared All-Sky Survey
(IRAS) archive and images at 8.8~$\mu$m from the Midcourse Space Experiment
(MSX) archive. Finally we use \mbox{X-ray} data from the {\it ROSAT} All-Sky
Survey (RASS) Faint Source Catalog \citep{Voges2000} to complete our
multi-wavelength data set.

\section{Analysis and Results}

\subsection{SNR~B0513--692}

\subsubsection{Radio}

\begin{figure}
\begin{center}
  \includegraphics[scale=0.41, angle=0]{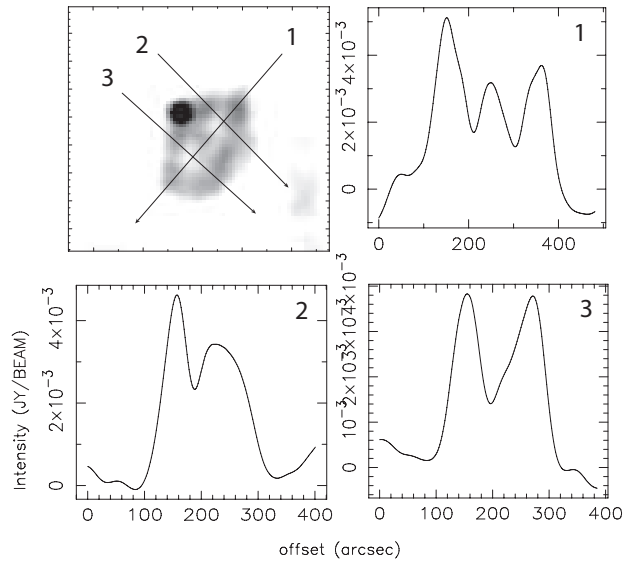}
\caption{Intensity slices through SNR~B0513--692.  The direction
of the arrows point from left to right from the zero offset on the
x-axis of the accompanying slices 1, 2 \& 3.
 \label{Fig2}}
\end{center}
\end{figure}

The ATCA 4790~MHz radio image (Fig.~\ref{Fig1} top) of
SNR~B0513--692 shows an elliptical ring-like structure with central
features consistent with the new 1377~MHz ATCA mosaic image (Staveley-Smith et al., in prep.). The SW
part of the elliptical shell is reasonably symmetrical with peak
fluxes including values of 3.65~mJy Beam$^{-1}$ (SE contour at
7.3$\sigma$),~4.4~mJy Beam$^{-1}$ (SW contour at 8.8$\sigma$) and
5.15~mJy Beam$^{-1}$ (NW contour at 10.3$\sigma$). The structure of
the northeast half of the shell cannot be resolved since the strong,
point-like source, J051324--691049, is embedded towards the NE edge of the SNR.
Two cavities (which are unresolved - they match the beam size) are
located in the central part of the SNR with fluxes below the
4$\sigma$ level. In Fig.~\ref{Fig2} we show the relative intensities
of slices drawn through these central cavities. Slices along the
minor axis show a slightly steeper gradient on the north eastern
side of the shell which may be explained by a gradient in the
surrounding ISM density or magnetic field. The shell thickness in the radio
is measured at approximately 30\% of the objects radius, which is probably just
an upper limit considering the size of the restoring beam used.

The remnant shell in our ATCA 8640~MHz image (Fig.~\ref{Fig1} bottom) was
restored using a 34$\times$34~arcsec gaussian beam and has a
very low dynamic range above the surrounding noise. The unreliable
contour at 2$\sigma$ shows basic similarities to the 3$\sigma$
contour of the 4790~MHz image, but some features including the
southwestern ridge are inconsistent. These regions were most likely
poorly reconstructed during the cleaning stage, resulting in the
formation of `blobs' on the image.

The major and the minor axes are estimated
from Fig.~\ref{Fig2} to be 260$\times$170~arcsec in angular extent.
Given an LMC distance of 51.5~kpc \citep{Feast1999,Nelson2000}, this gives
the linear size of the remnant as 65$\times$42~pc ($\pm7$ pc for
both axes). By comparison, optical AAT imaging in H$\alpha$ \citep{Math1985} give
dimensions of 60$\times$51~pc (using a distance to the LMC of 55 kpc).

\begin{figure}
\begin{center}
 \includegraphics[scale=0.35, angle=0]{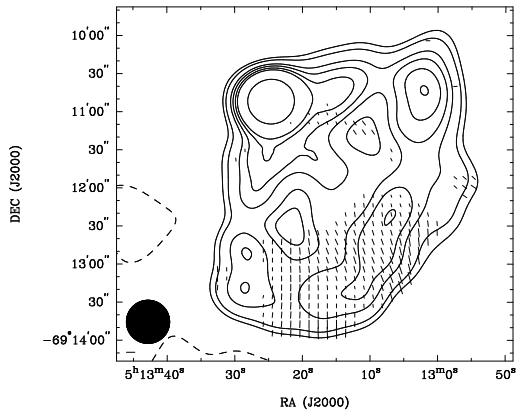}
\caption{Linear polarization image of SNR~B0513--692 at 4790~MHz
with the same intensity contours used in Fig.~\ref{Fig1}.
Polarization vectors are drawn along a band in the middle of the
remnant as well as in its southern hemisphere. The length of these
vectors are proportional to the fractional polarization, with the
highest noted at 49\%. This amount of fractional polarization is
denoted by the bar in the lower left corner of the image. Note that
the correction for the Faraday rotation could not be applied (see
text).
 \label{Fig3}}
\end{center}
\end{figure}

Linear polarization images for each frequency were created using
\textit{Q} and \textit{U} parameters. While we detect no reliable
polarization at 8640~MHz, the 4790~MHz image reveals strong linear
polarization in a band along the center of the remnant and its
southern hemisphere along the rim. Fig.~\ref{Fig3} shows areas of
fractional polarization superimposed on the same 4790~MHz intensity
contours used in Fig.~\ref{Fig1}. It is the first reliable detection of
polarized radio emission from this object. Polarization vectors were plotted
only in those regions where the polarized intensity was greater than
$3\sigma_{q,u}$ above the noise level for the \textit{Q} and
\textit{U} maps ($\sigma_{q,u} = 0.1\textrm{ mJy Beam}^{-1}$). The
length of these vectors are proportional to the amount of fractional
polarization with a maximum length corresponding to 49\%. The
relatively small amount of linearly polarized emission on the one
hand and the patchy and somewhat dubious total intensity image on
the other, prevent us detecting any significant polarization at
8640~MHz (beyond instrumentation error). Without reliable
polarization measurements at the second frequency we could not
determine if any Faraday rotation was present.

The mean fractional polarization at 4790~MHz was calculated using
flux density and polarization:
\begin{equation}
P=\frac{\sqrt{S_{Q}^{2}+S_{U}^{2}}}{S_{I}}\cdot 100\%
\end{equation}

\noindent where $S_{Q}, S_{U}$ and $S_{I}$ are integrated
intensities for \textit{Q}, \textit{U} and \textit{I} Stokes
parameters. Our estimated value is $P\cong 10\%$.

\begin{table*}
\begin{center}
 \caption{Integrated flux densities for SNR~B0513--692 and the compact source
 J051324--691049. Columns 5 and 6 list flux densities for high and low
 resolution images.
 \label{Tab2}}
\small{\begin{tabular}{cccccc} \hline \noalign{\smallskip}
(1)&(2)&(3)&(4)&(5)&(6)\\
Frequency & Noise$_{\mathrm{rms}}$ & B0513--692 & J051324--691049 & J051324--691049 & J051324--691049\\
    &      &  $S_{\nu}$   &     $S_{\mathrm{Peak}}$
&   $S_{\nu}$ (high res.) & $S_{\nu}$ (low res.) \\
$[\textrm{MHz}]$  &  $[\textrm{mJy Beam}^{-1}]$ & [mJy]  &
$[\textrm{mJy Beam}^{-1}]$    &  [mJy]  &   [mJy]\\
\noalign{\smallskip} \hline \noalign{\smallskip}
  \p0408   &         ---   &       360.0  &         ---     &         ---    &           --- \\
  \p0843   &         ---   &       302.0  &         28.6    &         ---    &           33.0\\
   1377    &           0.5 &       260.0  &         24.8    &         ---    &           25.0\\
   4790    &           0.5 &     \p087.0  &         15.4    &         16.0   &           25.4\\
   8640    &           0.4 &     \p030.0  &         \p08.4  &       \p09.0   &           19.9\\
\hline \noalign{\smallskip}
\end{tabular}}
\end{center}
\end{table*}

\begin{figure}
\begin{center}
\includegraphics[scale=0.46, angle=0]{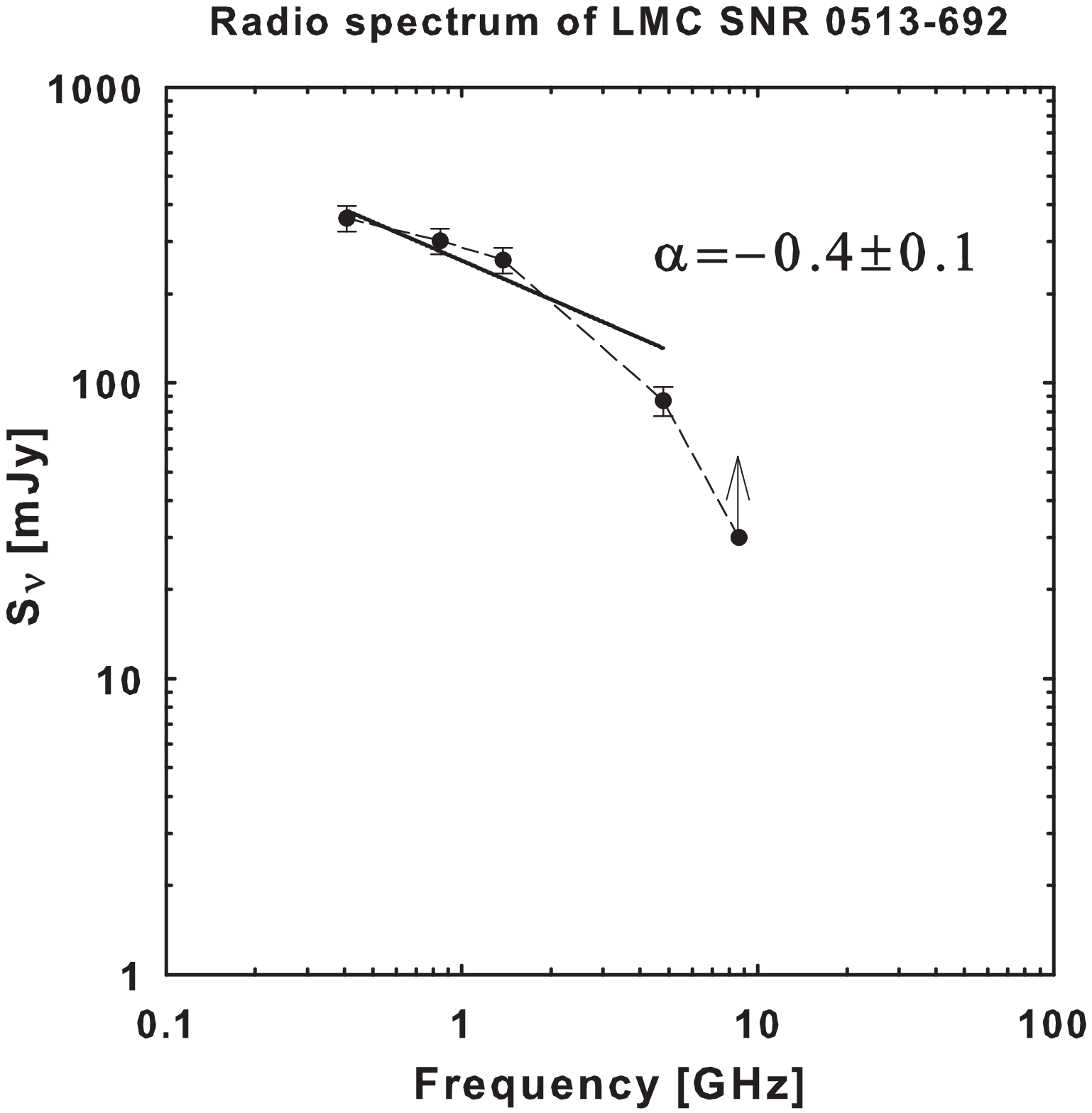}
\includegraphics[scale=0.57, angle=0]{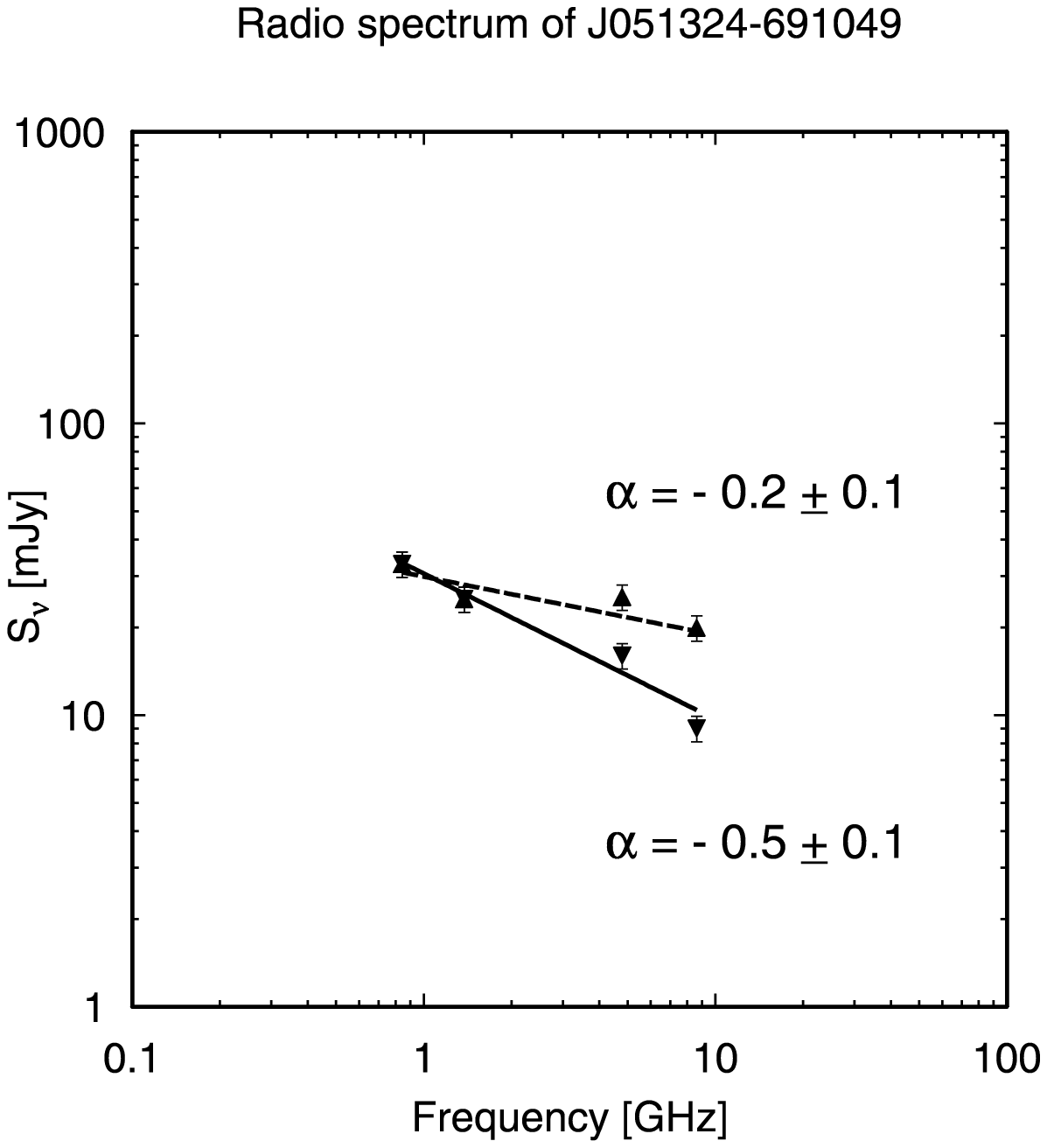}
\caption{Log-log graphs of the radio spectrum from
SNR~B0513--692 (top) and the embedded compact radio source J051324--691049
(bottom). The flux density of SNR~B0513--692 at 8640~MHz is most
likely a lower limit of the actual value and was excluded from the
power-law fit. It is more likely that the spectrum of SNR~B0513--692
is curved as indicated by the dashed line.
No missing flux was detected at higher frequencies (8640/4790~MHz)
after comparison with lower resolution Parkes images. The spectral
analysis of J051324--691049 included flux values from the high resolution
image (dashed line) and the low resolution image (solid line).
\label{Fig4}}
\end{center}
\end{figure}

These results indicate the presence of a relatively strong and well
organized magnetic field with some kind of radial structure
(considering the fact that the observed strong polarization
intensity match parts of the shell where the total intensity
emission is also strong).

By fitting an unweighted regression line (i.e. power-law fit) at 4 frequencies
using the flux density from our 4790~MHz observations, together with 408, 843
(MOST) and 1377~MHz (ATCA-mosaic) flux densities (Fig.~\ref{Fig4}),
we obtained a spectral index of $-0.4\pm0.1$ for SNR~B0513--692. This
confirms the objects non-thermal nature.
Integrated flux densities and rms noise for each of these
wavelengths are listed in Table~\ref{Tab2}. Our flux densities at 4790 and
particularly 8640~MHz most likely represent lower limits because of the relatively
short integration time. The flux density at 8640~MHz was omitted from the regression-line calculation for
this reason.

Flux density errors ($\sigma_s$) were estimated from: \mbox{$\sigma_s=[N\cdot$Noise${_{rms}}^2+(0.05\cdot
S_{\nu})^2]^{1/2}$} \citep{Weiler1986, HoUl2001}, i.e. from the rms noise across the surface of a source,
and from the influence of the errors in calibration. Since the calculated values range between 6\% and 10\%,
we assign 10\% error bars to the remainder of the flux as an upper limit of flux density errors.

\subsubsection{Previous \mbox{X-ray} Observations}

This SNR was first detected in
the Einstein survey of the LMC \citep{Long1981} and
diffuse \mbox{X-ray} emission from SNR~B0513--692 was
seen in the Rosat All Sky Survey (RASS) data of the region and also
in the Position Sensitive Proportional Counter (PSPC) images.
It is listed in the RASS Faint Source
Catalogue as 1RXS~J051315.7--691219 \citep{Voges2000} and in the LMC PSPC
catalogue by \cite{haberl1999} as HP~835. Table~\ref{Tab3} cites additional
information taken from this catalogue.

A useful method to characterize the spectrum of \mbox{X-ray} sources
with limited counts is through the calculation of hardness ratios.
These (HR1 and HR2) are defined by taking the ratio of the
differences in count rates between a set of \mbox{X-ray} energy
bands:
\begin{equation}
\textrm{HR1}=\frac{(hard-soft)}{(hard+soft)}
\end{equation}
and
\begin{equation}
\textrm{HR2}=\frac{(hard2-hard1)}{(hard2+hard1)}
\end{equation}
\noindent In this case, the energy bands are defined as: $soft$
(0.1--0.4~keV), $hard$ (0.5--2.0~keV), $hard1$ (0.5--0.9~keV) and
$hard2$ (0.9--2.0 keV). SNRs in the LMC have a wide distribution of
HR2 values between -1 and +0.55 \citep{Fica1998}. The HR2 value for
SNR B0513-692 of -0.34$\pm$0.17 (Table~\ref{Tab3}) is situated very
close to the peak of this distribution.

\begin{table}
\begin{center}
\caption{SNR J0513--692 X-ray data from the {\it
ROSAT} PSPC survey. \label{Tab3}} \small{\begin{tabular}{ll} \hline
\noalign{\smallskip}
RA(J2000)        &  $  05^{h}13^{m}13.5^{s}$   \\
DEC(J2000)     &   -69\D 11\arcmin 30\arcsec     \\
Total Positional Error&  38.7~arcsec            \\
Count Rate [counts s$^{-1}$ ]     &  $1.41~(\pm0.28)\times 10^{-2}$ \\
HR1                 &  $+1.00\pm0.42 $  \\
HR2                 &  $-0.34\pm0.17 $    \\
\hline
\end{tabular}}
\end{center}
\end{table}

\subsubsection{Previous and new Optical Observations}

In order to complete our multi-frequency discussion of
SNR~B0513--692, we discuss the previous optical observations of
this remnant in more detail. Optical counterparts for
SNR~B0513--692 have previously been detected in the \HA,
\SII\ and \OIII\ emission lines from AAT narrow-band IPCS image data from \cite{Math1985}.

\begin{figure}
\begin{center}
\includegraphics[scale=0.35, angle=-90]{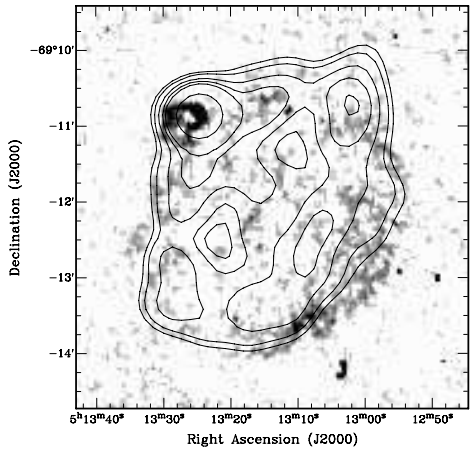}
\caption{Red continuum subtracted 5$\times$5 arcmin
\SII\ image of B0513--692 superimposed with the same 4790~MHz
radio contours used in Fig.~\ref{Fig1}.
(MCELS \SII\ and red images provided courtesy of C. Smith). \label{Fig5a}}
\end{center}
\end{figure}

We show in Fig.~\ref{Fig5a} the new \SII\ image of the remnant from the
deep MCELS data of the LMC. An $5\times5$~arcmin area extracted
around the position of the radio position of the SNR reveals a
coherent, oval shell structure with internal filamentary features
but with a strong "hint" of a possible mirror symmetry with
two regions with low brightness lying on the major axis.
The major and minor axes dimensions are $245\times200$~arcsec
equating to $61\times50$~pc at the distance to the LMC. This
grey-scale \SII\ image is also overlaid with the 4790~MHz radio
contours of Fig.~\ref{Fig1} which are seen to be well matched in
overall extent. This strong optical image from the \SII\
6717/6731\AA\ emission doublet is indicative of shock-excited gas
and a useful diagnostic for SNR optical detection, especially when
the \SII\ point to point emission line strength is $>0.5$ times that
for \HA\ (e.g. \cite{Fesen1985}) as already found by \cite{Math1985}.

\begin{figure}
\begin{center}
\includegraphics[scale=0.35, angle=-90]{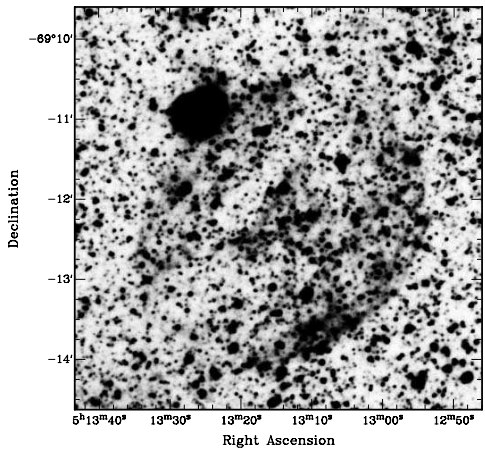}
\caption{A $5\times5$~arcmin \HA\ divided by short-red
(continuum) image of SNR B0513--692 taken from the deep Reid/Parker \HA\ and SR
median stacked images of the central 25sq.deg of the LMC (Reid \& Parker 2006a,b).
Note the strong similarity with the equivalent \SII\ image from MCELS (Fig.~\ref{Fig5a})
though the compact, strong emission in the NE edge of the SNR is saturated in this presentation,
preventing any discernment of internal structure there. \label{Fig5b}}
\end{center}
\end{figure}

The equivalent \HA\ image of the SNR taken from the new deep
H-alpha/SR median stacked image of the central 25~sq.deg of the
LMC of Reid/Parker (Reid \& Parker 2006a,b) is shown in
Fig.~\ref{Fig5b}, obtained by dividing the \HA\ image by the
matching short-red continuum image. This new \HA\ map has allowed
extremely low surface brightness nebulosities and structure to be
detected in the vicinity of B0513-692. This is due its arcsecond resolution
and high sensitivity ($R_{equiv}\sim22$ for \HA\ or
$4.5\times10^{-15}ergs~cm^2~s^{-1}$ \AA$^{-1}$). The \HA\ map is
strikingly similar to the \SII\ map with all the same, major filamentary
features evident within the overall oval structure. In \HA\ the SNR
appears to be a clear example of tilted barrel-shaped morphology
\citep{KestCasw1987, Gaensler1998} with the symmetry axis passing through the
bright central filament.

\begin{figure}
\begin{center}
\includegraphics[scale=0.46]{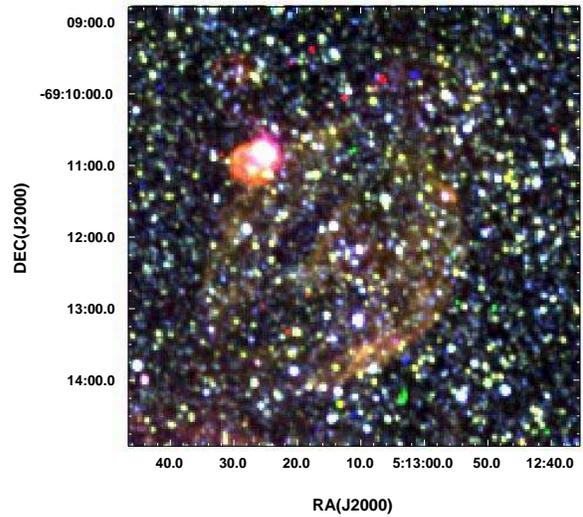}
\includegraphics[scale=0.46]{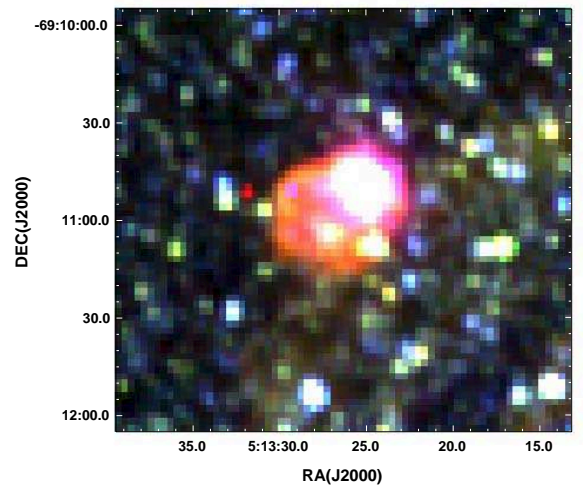}
\caption{Pseudocolour representations of SNR B0513--692 and
J051327-6911. The images are a straight combination of the \HA~(red), \SII~(green) and
\OIII~(blue) images obtained by MCELS (courtesy C. Smith). The overall
optical morphology and angular size of B0513--692 (top) is in good
agreement with the new radio images at 3 and 6~cm. The bottom image gives the enlarged
NE part of the top image. The bright, orange circular object on the periphery of B0513--692 is
clearly distinct both from the large oval shell of B0513-692 itself and the compact, pink-white
HII region immediately to the NW. \label{composite}}
\end{center}
\end{figure}

Finally, a new MCELS colour composite \HA, \OIII\ and \SII\ image
is presented in Fig.~\ref{composite}. This image clearly brings out the
nature of SNR~B0513--692 and further details concerning the
compact region on the North-Eastern limb (see section~\ref{small_snr} below).


\subsection {Commpact radio source J051324--691049 (optical counterparts LHA 120--N112; DEM\,L109)
 \label{small_snr}}

In the North-East edge of B0513-692 there is the compact radio
source J051324--691049. Previous radio observations of
J051324--691049 suggested it was a compact \HII\ region
\citep{Math1985}. An optically identified object at this position
was also listed in the \cite{Henize1956} catalogue of Magellanic
Clouds emission nebulae as LHA~120--N\,112 based on low dispersion
objective prism photographic spectroscopy and as DEM\,L109 in the
catalogue of nebular complexes of the Large and Small Magellanic
Clouds by \citep{Davies}.

The SIMBAD database identifiers associated with J051324--691049 are
listed in Table~\ref{Tab4} and we consider them all as refering to
the same source. However, \cite{Gurwell} listed a galaxy behind the
LMC with a position coincident with J051324--691049 based on CTIO
4-m broad-band optical B \& V photography. They listed coordinates
(J2000) of RA=05$^{h}$13.4$^{m}$ and DEC=--69\D 11\arcmin, citing
rather large positional uncertainties of about $\pm10^{s}$ in right
ascension and $\pm1$ to 2~arcmin in declination.
Unfortunately, this led to a confusing connection in the NASA/IPAC
Extragalactic Database (NED) of radio, far and near-infrared
coincident sources with this putative galaxy. Also note that the
coordinates of N\,112 listed by SIMBAD in Table~\ref{Tab4} are
incorrect. Here, we provide updated, accurate, optical co-ordinates
and attempt to clarify the true nature of the compact N\,112 region
and its eastern extension through an analysis of new radio,
infrared, CO and optical data.

\subsubsection{Radio}

The compact radio source, J051324--691049 is embedded within the
North-East edge of \mbox{SNR~B0513--692}. It has a peak flux
position at RA=05$^{h}$13$^{m}$24.8$^{s}$ and \mbox{DEC=--69\D
10\arcmin 49.1\arcsec} (J2000.0). A radio spectral index was
constructed using flux densities obtained from our 8640 and 4790~MHz
ATCA images and data from other radio frequencies as shown in
Table~\ref{Tab2}. Column 4 lists peak fluxes, column 5 flux
densities from our high resolution images (using all antennas) and
column~6 flux densities from our low resolution images (excluding
antenna 6). We use the {\sc miriad} task {\sc imfit} to determine
position and peak/integrated flux values for our data. Flux density
values for 843 and 1377~MHz data are the same in both columns~5 and
6.

\begin{table*}
\begin{center}
 \caption{Identifiers from the SIMBAD database giving previously catalogued objects
 in the immediate vicinity of the point-like radio source J051324--691049.
 \label{Tab4}}
\small{\begin{tabular}{llll} \hline \noalign{\smallskip}
Catalogue Name & RA(J2000) & DEC(J2000) & Reference\\
\hline \noalign{\smallskip}
J051324--691049&05$^{h}$ 13$^{m}$ 24.8$^{s}$&-69\D 10\arcmin 49.1\arcsec&This paper\\
\hline
LHA 120--N\,112 &05$^{h}$ 13$^{m}$ 23.2$^{s}$ &-69\D 11\arcmin 38.1\arcsec&\cite{Henize1956} \\
LHA 120--N\,112 &05$^{h}$ 13$^{m}$ 14.6$^{s}$ &-69\D 13\arcmin 37.0\arcsec&SIMBAD\\
DEM L109        &05$^{h}$ 13$^{m}$ 26.5$^{s}$ &-69\D 10\arcmin 53.3\arcsec& \cite{Davies}\\
GH 6--2         &05$^{h}$ 13$^{m}$ 24.0$^{s}$ &-69\D 11\arcmin            &\cite{Gurwell} \\
IRAS 05137--6914&05$^{h}$ 13$^{m}$ 24.67$^{s}$&-69\D 10\arcmin 48.4\arcsec&Joint {\it IRAS} Working Group\\
MSX LMC 217     &05$^{h}$ 13$^{m}$ 24.67$^{s}$&-69\D 10\arcmin 48.4\arcsec&\cite{Egan2003}\\
No. 180         &05$^{h}$ 13$^{m}$ 18.8$^{s}$ &-69\D 10\arcmin 38\arcsec   &Kawamura et al. in prep\\
OGLE-CL LMC241  &05$^{h}$ 13$^{m}$ 25.65$^{s}$&-69\D 10\arcmin 50.1\arcsec&\cite{pietr99}\\
\hline
\end{tabular}}
\medskip\\
\end{center}
\end{table*}

If we assume that all of the radio flux from J051324--691049 is not
fully resolved in the high resolution imaging process, then the
spectral index, fitted from the integrated flux densities (sixth
column in Table~\ref{Tab2}), is relatively flat with
\mbox{$\alpha=-0.2\pm0.1$} as shown by the dashed line in Fig.~\ref{Fig4}.
On the other hand, is is possible
that this radio source is unresolved (i.e. if its angular
dimensions are less than 1~arcsec beamwidth of the restoring beam
used in deconvolution). In this case the spectral index would be strongly
non-thermal with \mbox{$\alpha=-0.5\pm0.1$} as
shown by the solid regression line in Fig.~\ref{Fig4}.

We have calculated the brightness temperature for this compact radio
source at 8640 and 4790~MHz using the equation:
\begin{equation}
\textrm{T}_B=\frac{\lambda^{2} S_{\lambda} }{2k \Omega_{\lambda}}
\end{equation}
where $S_{\lambda}$ is flux density and $\Omega_{\lambda}$ is
angular size in steradians. The resultant value for $\lambda\simeq$~3.5~cm
is T$_{B}^{3.5cm}=148~$K and for $\lambda\simeq$~6~cm is
T$_{B}^{6cm}=213~$K. Since our flux densities at both frequencies
represent lower limits, these numbers are also lower limits of
brightness temperature. They do not rule out the presence of thermal
radio emission from this object.

The polarization estimates for J051324--691049 at each wavelength were found to be
below the noise level (i.e. the upper limit for polarization is
\mbox{$\le$0.6 mJy Beam$^{-1}$} and $\le0.3$~mJy~Beam$^{-1}$ for
8640 and 4790~MHz, respectively) and thus not helpful in
characterizing radio emission from this object.

\subsubsection{CO detection as a proxy for molecular hydrogen}

We used data from the LMC CO survey undertaken with the 4-metre NANTEN millimetre-submillimetre radio
telescope\footnote{operated by Nagoya University at Las Campanas
Observatory in Chile and under mutual agreement with the Carnegie
Institution of Washington} (Kawamura
et al. in prep). One of the main aims of the NANTEN project is
the full-mapping of molecular clouds in the LMC and along
the Galactic plane using the $^{12}$CO$(1-0)$ transition as a proxy for molecular hydrogen.
A relatively weak molecular cloud, listed as
No.~180 in the LMC CO cloud catalogue was found in the vicinity of J051324--691049.
This small molecular cloud (compared to the NANTEN beam of
$\sim$2.7~arcmin or 39~pc at the LMC distance) was found at (J2000.0):
RA=05$^{h}$13$^{m}$17.8$^{s}$ and DEC=--69\D 10\arcmin 38\arcsec.
The spectrum of the CO emission of this cloud has the absolute
antenna temperature, Tr*=0.49~K, with LSR velocity
V$_{lsr}$=233.4~km/s, and a FWHM velocity width of
$\Delta$V=2.34~km/s. For the LMC NANTEN survey, they take a
velocity-integrated intensity, II=1.2~Kkm/s on average, as
representing a real detection (3~$\sigma$ noise level). We find only
one spectrum toward this region which satisfies the above criteria,
though we can `see' the spectrum adjacent to it. If we take a
2~arcmin grid spacing and define a radius `R' from the detected
cloud area of $\pi \times$ r$^2$, then we find R=16.41~pc (assuming
D=51.5~kpc). This yields a virial mass
M$_{vir}$=1.7$\times$10$^{4}$~M$_{\odot}$. The observed line-width
of 2.34~km/s is one of the narrowest seen among CO emissions in the
LMC detected by NANTEN. The average $\Delta$V is $\sim$6~km/s.

We found that a significant number ($>$75\%) of known LMC SNRs appear to be
associated with CO clouds (Kawamura 2006, priv. comm.). This is
evidence that the SNR shock front is interacting with the
surrounding environment. It is reasonable to
expect that most SNRs are located around dense molecular clouds.

However, the question remains as to whether this molecular cloud is
physically connected with B0513--692 or indeed associated with
another source given the positional uncertainties of
$\sim$2.7~arcmin associated with the CO maps. Only higher resolution
CO observations could resolve this question.

\subsubsection{Infrared and Mid-Infrared emission}

An entry in the {\it IRAS} Point Source Catalogue, version 2.0 (NASA RP--1190):
IRAS~B05137 --6914 gave a decent positional match to J051324--691049.
Its flux densities in the different IRAS bands together with quality measurements, are
listed in Table~\ref{Tab5}. We also note that the {\it IRAS}
observations gave no significant infrared emission from SNR~B0513--692.

\begin{figure}
\begin{center}
\includegraphics[scale=0.4, angle=0]{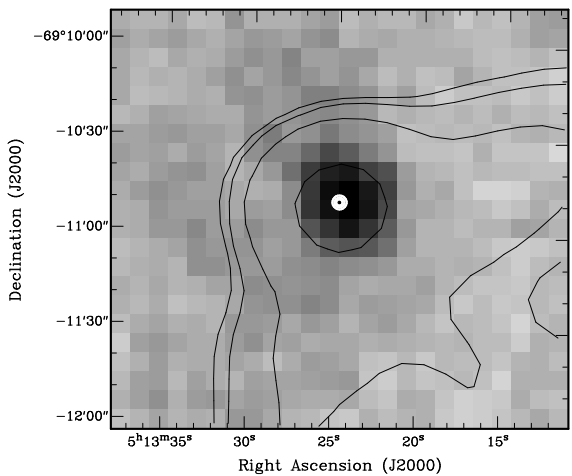}
  \caption{MSX 8.3~$\mu$m grey-scale image of J051324--691049
        overlaid with 4790~MHz low resolution (black) contours
        (3, 4, 7 and 25$\sigma$; $\sigma=0.5~\textrm{mJy Beam}^{-1}$)
        and high resolution (white) contours (4, 6 and 8~mJy).
 \label{Fig6}}
\end{center}
\end{figure}

In Fig.~\ref{Fig6}, we show an~8.3~$\mu$m (Band-A) high
sensitivity image of J051324--691049 from MSX where the astrometric accuracy permits
a much more unambiguous match. The source is seen in all 4 MSX bands
but it is strongest at~8.3~$\mu$m. Flux densities from all four
bands are listed in Table~\ref{Tab5} (\citealt{Egan2003}). This ~8.3~$\mu$m
MSX image reveals a compact, centre-brightened, but resolved
source with an elliptical shape. Its angular dimensions are
$\sim$~30$\times$40~arcsec (7.5$\times$10~pc) and its
position, with positional uncertainty of
$\sim1$~arcsec, is presented in Table~\ref{Tab4}. The same
source is catalogued as record number 20 in the catalogue of
cross-correlated LMC sources from MSX and The Two Micron All Sky
Survey (2MASS) surveys, \cite{Egan}. Data from matching sources
from those two surveys are presented in
order to expand the IR colour baseline. Furthermore, considering
the much improved combination of sensitivity and resolution of
MSX compared to IRAS, better insight into
LMC objects with high IR excess (AGB stars, planetary nebulae
(PNe) and compact \HII\ regions) is expected. The 2MASS $J$, $H$,
$K$ survey photometry measured at the J051324--691049 position are
also presented in Table~\ref{Tab5}. The resultant IR colours of the
source with $J-K_s<2$, $K_s-$A$\geq2.5$ and $0.75\leq H-K_s<2$,
J051324--691049 are typical of PNe but also overlap with those of
\HII\ regions (\cite{Egan}). We can resolve the ambiguity by
simple consideration of the angular extent and hence physical size of
J051324--691049 which is estimated at $\sim10$~pc, far too large
for any PN. This makes a compact \HII\ region the most probable
identification. It is most likely powered by the dominant compact stellar
association located in the centre of the region which can be seen
clearly in the SuperCOSMOS B-band image. This stellar content of
the \HII\ region is catalogued as the star cluster OGLE-CL LMC241
\citep{pietr99}.

\begin{table}
\begin{center}
 \caption{Near-IR and mid-IR flux densities and magnitudes of IRAS 0513-6914
from the {\it MSX, 2MASS} and {\it IRAS} surveys.
 \label{Tab5}}
\small{\begin{tabular}{clllr} \hline \noalign{\smallskip}
Band & Wavelength & Int. Flux & mag & Survey\\
 & ($\mu$m) & (Jy) & &  \\
\hline
-&12.0 & 0.7824 &&{\it IRAS}\\
-&25.0 & 2.584 &&{\it IRAS}\\
-&60.0 & 19.71 &&{\it IRAS}\\
-&100.0 & 58.1 &&{\it IRAS}\\
A&8.28 & 0.33$\pm$0.01 & 5.92 &MSX\\
C&12.13 & 0.50$\pm$0.05 &&MSX\\
D&14.65 & 0.53$\pm$0.04 &&MSX\\
E&21.34 & 1.10$\pm$0.08 &&MSX\\
$J$&1.0-1.5 & & 14.7$\pm$0.1 &2MASS\\
$H$&1.5-2.0 & &14.21$\pm$0.1 &2MASS\\
$K_s$&2.0-3.0 & & 13.23$\pm$0.1 &2MASS\\

\hline
\end{tabular}}
\medskip\\
\end{center}
\end{table}

\subsubsection{Narrow-band optical emission-line imaging}

To further characterize the nature of the region around the
compact radio source we closely examined the new, deep \SII, \OIII\ and
\HA\ images of J051324--691049.

First, we re-examined the red-continuum subtracted MCELS \SII\
image of B0513--692 shown in Fig.~\ref{Fig5a}.
There is a strong, compact, \SII\ emission
feature on the North-East rim of the SNR that coincides with
the compact source J051324--691049. Interestingly, however,
there is an additional, quite strong, shell like feature immediately
to the East that blends in to the compact source but appears
somewhat distinct compared to the larger, oval structure associated
with SNR B0513-692 in terms of emission line strength and shape.

We have also re-examined data from the \HA\ map of the LMC referred
to in section~1 (see Reid \& Parker 2006a,b for further details)
together with the equivalent matching broad-band red `SR' image.
This matching \HA\ image in  Fig.~\ref{Fig5b} has been manipulated
to highlight the low surface-brightness coherent details across the
whole SNR. Note the region around the compact \HII\ region N\,112 is
completely saturated in this representation but hints of a two
component nature to the compact saturated zone is seen.

A small extract of the \HA\ grey-scale image is shown in
Fig.~\ref{Fig8}. Here, the data is shown at base contrast with a
linear pixel intensity scale. The compact \HII\ region
J051324--691049 is clearly seen as an intense, approximately
circular structure about 20~arcsec across but with an irregular
border. However, the raw, \HA\ continuum subtracted image also
reveals an adjacent, fainter, approximately circular nebula
structure $\sim40$~arcsec across and about 15~arcsec to the
South-East. We propose that this feature is not simply an extension
of the \HII\ region but is a separate source in its own right. This
proposition is neatly confirmed when we examine the MCELS combined
\HA\ (red), \OIII\ (blue) and \SII\ (green) image of the region as a colour composite
(Fig.~\ref{composite}~bottom). The ways the colours combine according to the relative intensity of
each emision line highlights the different nature
of the various components quite effectively. A clear, very well defined, circular, orange
nebula is seen immediately adjacent to the compact pink-white \HII\ region
and clearly brighter than and distinct from the fainter emission levels from the larger
scale B0513--692 remnant. The north-east part of B0513--692 intersects
with this object across $\sim50$\% of its area.

We designate this newly identified source as SNR~J051327--6911. This
new feature also matches the \SII\ extension to the compact source
seen in Fig.~\ref{Fig5a}.

Despite close angular proximity, the relative
\HA\ and \SII\ image intensities and morphologies of the compact
radio source J051324--691049 and our newly optically identified proposed SNR
J051327--6911, are sufficiently distinct to consider them as
separate entities. Unfortunately, the existing radio
observations of the region are completely dominated by the flux emanating
from the compact \HII\ region so that any independent signal from the
adjacent proposed SNR is lost. Further observations with longer integration
time and a suitable array configuration to resolve out the area could potentially reveal the
radio position and morphology of this object.

\subsection{Optical spectroscopy of the compact \HII\ region and new adjacent SNR candidate}

\begin{figure}
\begin{center}
\includegraphics[scale=0.36]{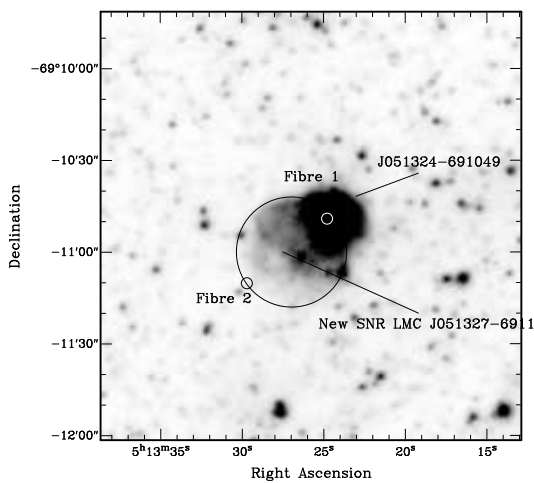}
\caption{AAO/UKST $12 \times 2$~hour exposure median-stack  \HA\
image (gray scale) of the N\,112 region (Reid \& Parker 2006a,b).
Two small circles marked as Fibre~1 (white) and Fibre~2 (black)
represent the positions of the two 2dF 2.5~arcsecond fibres placed
near the center of J051324--691049 (RA=05$^{h}$13$^{m}$24.8$^{s}$,
DEC=-69\D 10\arcmin 49.1\arcsec) and on the border of the newly suggested SNR
J051327--6911 (RA=05$^{h}$13$^{m}$29.71$^{s}$, DEC=-69\D 11\arcmin
19.2\arcsec) as far away as possible from N\,113.
The larger black circle is centered on the position
of the new proposed SNR J051327--6911 and it has an \Halpha\
diameter of $\sim$40~arcsec. Note that the \HA\ pixel intensities
of this coherent, small, shell-like structure adjacent to the
compact \HII\ region are twice that of the strongest optical
components of the large oval SNR which are not even visible in
this linear grey-scale image adding further support to its distinct nature.
 \label{Fig8}}
\end{center}
\end{figure}

\begin{figure}
\begin{center}
\includegraphics[scale=0.5, angle=0]{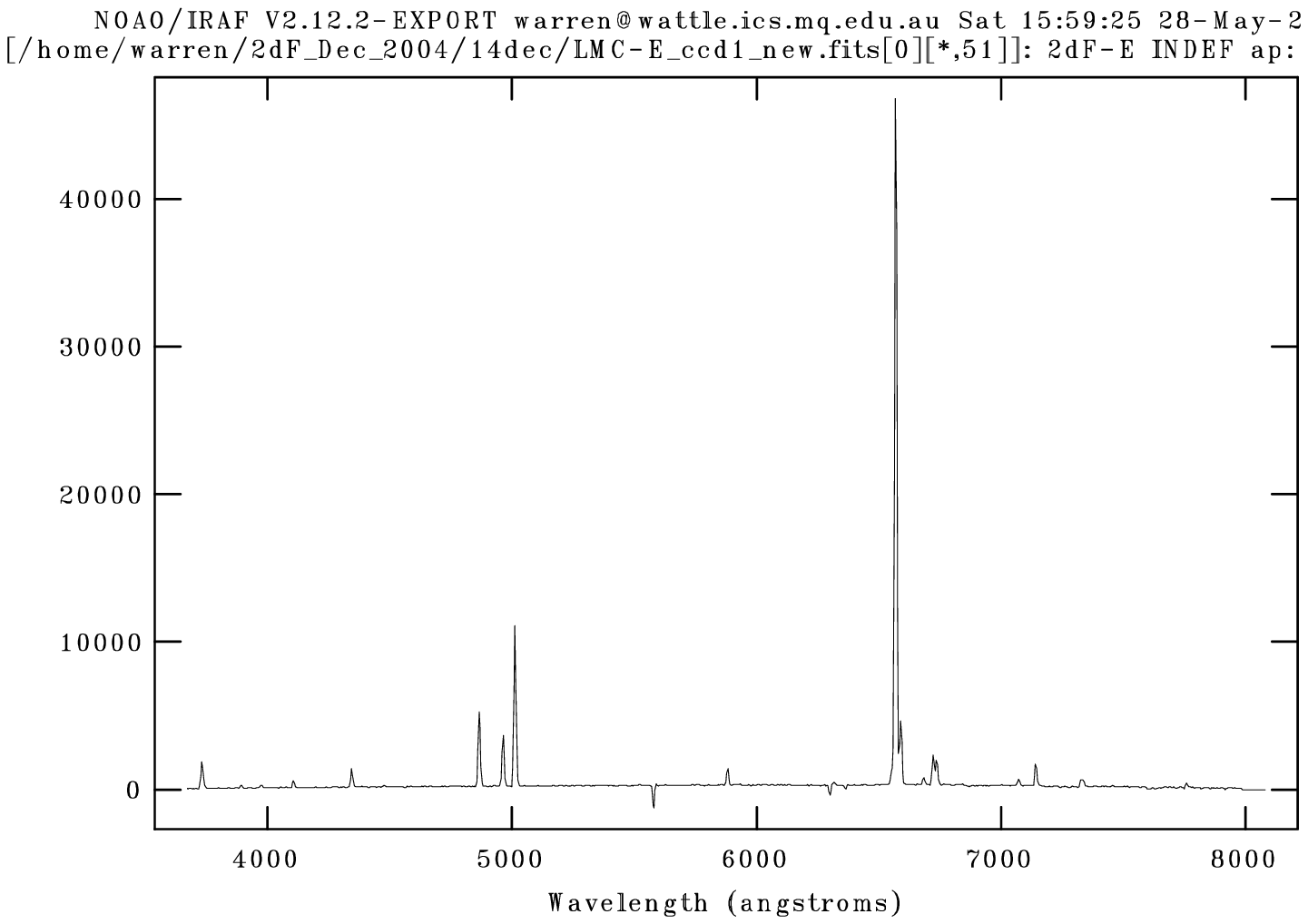}
\includegraphics[scale=0.5, angle=0]{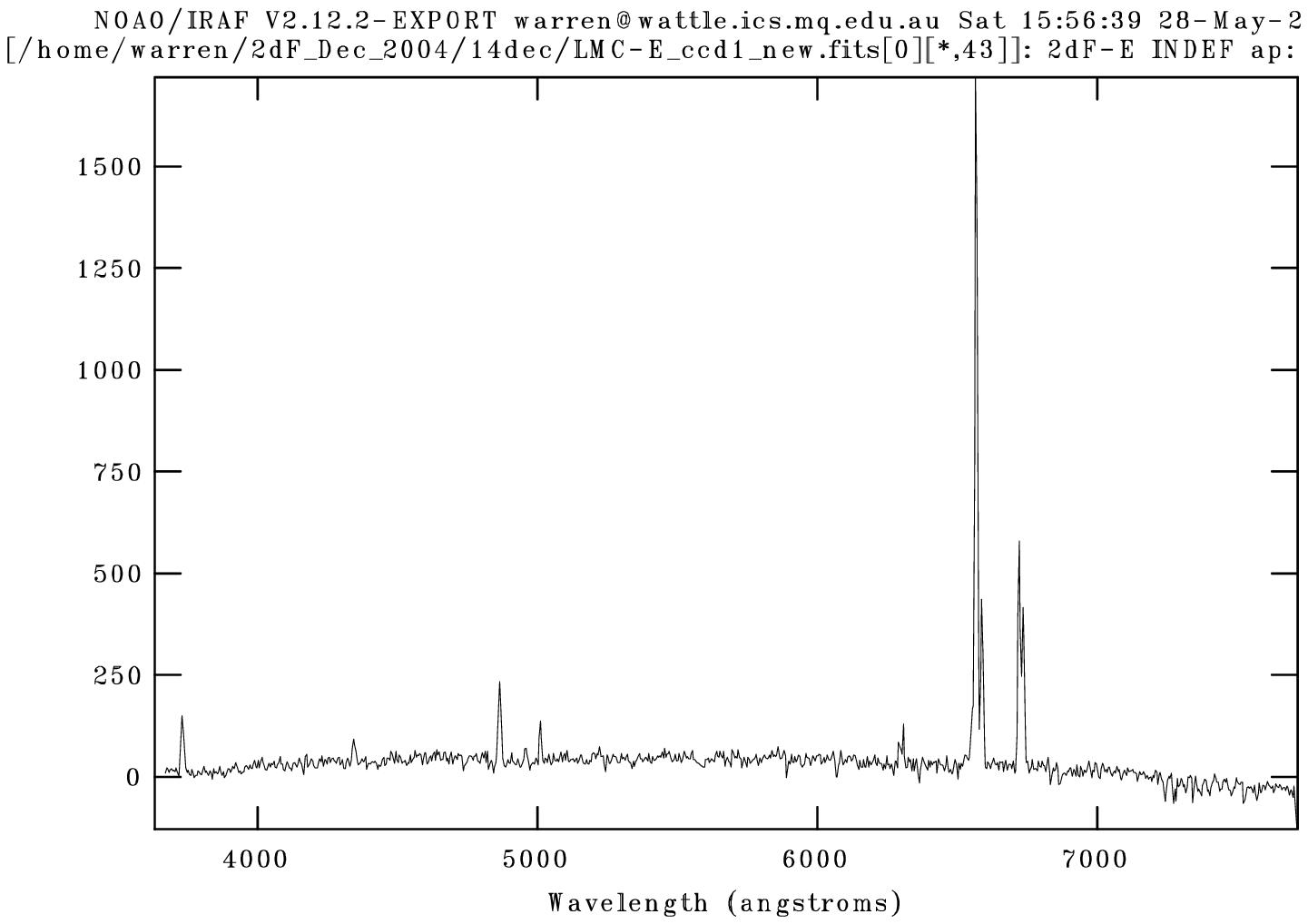}
  \caption{2dF reduced 1-D optical spectra of compact radio
        source J051324--691049 (top) and proposed new LMC SNR J051327--6911 (bottom).
        The top spectrum (Fibre~1; Fig.~\ref{Fig8}) is typical of a
        galactic \HII\ region whilst the lower spectrum (Fibre~2;
        Fig.~\ref{Fig8}) has strong \SII\ lines and prominent \OII3727\AA\
        and \OI6300, 6363\AA\ lines typical of an SNR.
 \label{Fig9}}
 \end{center}
\end{figure}

We obtained low resolution optical spectroscopy of the compact
radio source J051324--691049 and the newly suggested adjacent SNR~J051327--6911
located 15~arcsec to the south-east (Fig.~\ref{Fig8}). The spectra
were taken with the 2dF multi-object fibre spectroscopy system on
the Anglo-Australian Telescope (AAT) in December 2004
as part of an extensive follow-up programme of newly identified LMC emission sources,
\cite{ReiPar2006b}. The reduced, wavelength calibrated, sky-subtracted, 1-D
spectra of the two adjacent regions are given in Fig.~\ref{Fig9}. Note
that although no flux calibration was applied, the fibre relative
transmissions have been normalized via sky-line flux within the
reduction pipeline so that, given the identical exposure times and
observing conditions, the relative strengths of \HA\ and other
emission lines between the spectra are reasonably indicative. Note
that the \HII\ region \HA\ peak intensity is $\sim20\times$ that of
the adjacent source. The measured integrated intensities of the most
prominent lines for both objects as given from Gaussian fits are
given in Table~\ref{line_intensites}. As expected, based on the evidence
from the multi-wavelength images, the optical spectrum centred on the compact radio
source J051324--691049 (Fibre~1; Fig.~\ref{Fig8} and top panel of Fig.~\ref{Fig9}) is typical of a
\HII\ region or a very low excitation PN due to the absence of high
excitation lines and relatively weak \OIII\ relative to H$\beta$.
However, as noted previously, a PN is ruled out on the
basis of the large physical nebular size.
The spectrum of J051327--6911 (Fibre~2; Fig.~\ref{Fig8} and
bottom panel of Fig.~\ref{Fig9}) is completely different. The object
is clearly not what would be expected if it was from a faint
extension to the \HII\ region. Rather it is immediately suggestive
of an SNR due to the extremely strong \SII\ lines relative to
\Halpha\ and the prominence of the \OII3727\AA\ and \OI6300,
6363\AA\ emission lines. As shown in Table~\ref{line_ratios}, the
measured \SII/\HA\ ratio from this spectrum is 0.55 placing it fully within
the SNR domain following the line-ratio diagnostics of \cite{Fesen1985}.
This ratio is a prime indicator for distinguishing
SNRs from \HII\ regions, first pioneered for SNR searches in the
Magellanic Clouds by \cite{Math1973}. \HII\ regions typically show
\SII/H$\alpha$ ratios of about 0.1, while SNRs have ratios $\geq
0.4$ in most galaxies. These new spectroscopic data add
significant credence to the veracity of the new
SNR identification. The superposition of 2 SNRs which are likely at different stages of evolution,
is rare and provides interesting possibilities for further study.
The intersection along the line of
sight of SNR~B0513--692 and SNR~051324--691049 can be used to
test absorption of the closer remnant. With new, better quality
spectral line observation in the vicinity of this overlapping region
it may be possible to resolve remaining questions including the interrelation
of those two objects. Also, a new, much more sensitive radio observation
of these two SNRs could probe structure of the magnetic field in the
interlacing part of the shells through the measure of Faraday rotation.

Electron densities for these two sources calculated from the observed ratio
of the \SII\ lines are 385.2~cm$^{-3}$ and 37.2~cm$^{-3}$ for J051324--691049
and J051327--6911 retrospectively.

\begin{table}
\centering \caption{Measured line intensities of compact source
J051324-691049 (I$_{Fibre 1}$) and the new SNR candidate
J051327-6911 (I$_{Fibre 2}$).}
\label{line_intensites}
\begin{tabular}{llrr}
 \hline\hline \noalign{\smallskip}
$\lambda$~(\AA) &Line &I$_{Fibre 1}$&I$_{Fibre 2}$\\
 \hline \noalign{\smallskip}
3727&\OII&22605&1939\\
3835&H9&618&-\\
3869&\NeIII&548&-\\
3889&\HE&2153&-\\
3968&\NeIII&2571&-\\
4070&\SII&4586&-\\
4340&\HG&13940&947\\
4471&\HE&1419&-\\
4861&\HB&57374&2704\\
4959&\OIII&39732&625\\
5007&\OIII&125190&1004\\
5876&\HE&11952&-\\
6548&\NII&13849&980\\
6563&\HA&517888&18477\\
6584&\NII&46554&4360\\
6678&\HE&5198&-\\
6717&\SII&21016&5901\\
6731&\SII&19009&4329\\
7065&\HE&4584&-\\
7135&\AIII&15347&-\\
7323&\OII&9208&-\\
 \hline\noalign{\smallskip}
\end{tabular}
\end{table}

\begin{table}
\centering \caption{Line intensity ratios at Fibre~1 and Fibre~2
positions.}
\label{line_ratios}
\begin{tabular}{lll}
 \hline\hline \noalign{\smallskip}
Lines &Fibre~1&Fibre~2 \\
 \hline \noalign{\smallskip}
\NII/\HA&0.12&0.29\\
\SII/\HA&0.08&0.55\\
6717/6731&1.11&1.36\\
 \hline\noalign{\smallskip}
\end{tabular}
\end{table}

\section{Interaction between the components}

We found no evidence of any interaction between the two SNRs,
such as morphology deviation, or
enhanced \OIII\ and X-ray emission at the juncture of the possible
colliding region \citep{Will1997}. Such features are not apparent from the
currently available limited observational material. We think that direct
interaction is most unlikely since the young SNR appears just at the rim
of the older remnant which is a tight constraint on the position along
the line of sight. However, given the short relative lifetimes of
SNR the two remnants are likely to have exploded at a similar time and space.

There is a possible influence from high-energy UV photons
from the nearby star cluster OGLE-CL LMC241 to the ionization of
the young SNR, but no stratification of ionized material in the visible
part of the shell (outside of the \HII\ region) is detected.
With appropriate radio observations (wider ATCA configuration and longer
integration time), one could construct a distribution of the young SNR
spectral index, and map spectral changes across the remnant's shell. This
kind of distribution will show the influence and possible stratification of
the thermal emission (from photo-ionization) to the synchrotron radiation as
the main radio emission mechanism of an SNR.

According to \cite{Elada1977}, new generations of massive stars in OB
associations tends to occur in a sequential manner, i.e. star
formation moves away from the primordial cloud. Since the young SNR
appears centered well outside the nebula, one plausible scenario could be
that the progenitor of this remnant belongs to an older generation
of nearby massive stars. Perhaps some earlier SN explosions in an
earlier cluster triggered star formation in an adjacent molecular
cloud. So, all three objects (both SNRs and compact \HII\ region)
could have originated from the nearby molecular cloud as an
evolutionary effect.

\section{Discussion and Conclusions}

We verify B0513--692 as a new SNR but more importantly provide additional radio
data and new polarization measurements. Its polarized, steep ($\alpha$=-0.4$\pm$-0.1)
non-thermal radio emission and presence of \mbox{X-ray} emission, leave no doubt
of its true nature. The existence of a symmetry axis with features like mirror symmetry
and low-brightness regions at the end of the axis (see Fig.\ref{Fig5b}), place this
remnant in the barrel-shape morphological class \citep{KestCasw1987}.
Such a variations in the spherical nature of the remnant's structure could be
caused by a tube-like structure of the surrounding ISM \citep{Bisnovatyi1991}
or compression of an ambient magnetic field which generates high-brightness emission
regions in part of the shell where shock direction is perpendicular to the vectors
of the field \citep{Gaensler1998}. The later assumption is supported with the measured
presence of a relatively strong large-scale magnetic field in this region of the LMC
\citep{Gaensler2005}.

Presence of significant polarization
suggests that the magnetic fields within the shell are highly
ordered and relatively strong. In fact, we find that the estimated
level of polarisation of this SNR (49\%) is among the strongest ever
found for an SNR. There is no detected pulsar associated with SNR~B0513--692.
Available \mbox{X-ray} observations have insufficient spatial
resolution and sensitivity to detect an \mbox{X-ray} point source though
strong \mbox{X-ray} emission from the remnant was not expected due
to its large physical size \citep{Math1985}.

The embedded, compact, radio source J051324--691049, identified
previously as an \HII\ region (N\,112 or DEM\,L109) or background
source (GH~6-~2), is indeed confirmed as a compact \HII\
region, possibly powered by the dominant stellar association, OGLE-CL LMC241,
located at its centre. However, there is also an adjacent, faint optical
shell seen in both the MCELS \SII\ image and the deep, new
AAO/UKST \HA\ map of Reid \& Parker 2006a,b. It has a diameter of
$\sim$40~arcsec (10~pc), and is located about 15~arcsec to the
South-East of the compact radio source. We consider that this a
separate entity which we designate J051327--6911. The matching
\SII\ image reveals a strong source at this location compared to
the matching faint \HA\ emission, hinting at a likely SNR nature.
The subsequent optical spectroscopy for both J051324--691049 and
J051327--6911 exhibit markedly different spectral features with
the strong \SII\ relative to \HA\ for J051327--6911 being
indicative of shocked material and having a ratio typical of SNRs.
These optical observations combined with its symmetrical
structure, which shows a definite boundary, and different optical
image intensity, suggests that it is not a part of the more
extensive \mbox{SNR~B0513--692}. The compact and strong nature of
the radio source J051324--691049 most likely prevents any separate
signature from this adjacent SNR candidate being seen in the current radio data.

We strongly suggest that J051327--6911 is a new, separate, SNR.
Due to the small angular size of this new SNR candidate
one would assume a young nature but the lack of a confirmed strong
\mbox{X-ray} counterpart of this object could be a strong argument
against this presumption though we do propose a possible explanation for the unusually low \mbox{X-ray}
brightness of the source. To begin with the existing {\it ROSAT}
PSPC/HRI and Einstein observations are of low integration time and
therefore sensitivity. Furthermore, the observations did not cover
this area completely (the centre of the both ROSAT PSPC and HRI
pointings are well away from the new SNR candidate). Another
possibility is that this new SNR candidate is interacting with the
nearby high density molecular cloud and that the shock has already
decelerated to the point where the gas temperature downstream of the
shock is below the \mbox{X-ray} emitting temperature. This reduced
\mbox{X-ray} emission could be strongly absorbed along the line of sight
\citep{Ye1995}. Finally, the spatial resolution of previous X-ray
observations ($>$45~arcsec) would not be enough to resolve this new
and small SNR from the edge of larger nearby SNR B0513--692. Only
new Chandra/XMM and deeper ATCA observations of this area may solve
true nature of this intriguing object.

The objects are positioned in the Optical Bar and near the kinematical
center of the LMC. There are several \HII\ regions associated with
compact groups of young star clusters within $\sim150$~pc of the
observed field \citep{Yamaguchi2001} but nothing of the significance
and magnitude as the 30 Doradus or N\,11 star forming regions.

Finally the rare superposition of two SNRs in the LMC at different
stages of evolution, but presumably in a similar ISM environment,
provides interesting opportunities to unravel any possible interaction
and environmental issues through detailed chemical and kinematical analysis.

\section*{Acknowledgments}
We used the Karma software package developed by the ATNF and the
EXSAS/MIDAS software package developed by the MPE. We thank staff at
the MSSSO for the help during our optical observations. We also
thank Chris Smith for help in obtaining the MCELS images. Many
thanks to the anonymous referee for excellent suggestions.

\bsp

\label{lastpage}

\end{document}